\documentclass[10pt,a4paper]{article}
\usepackage[utf8]{inputenc}  
\usepackage[T2A]{fontenc}  

\usepackage{amsmath}
\usepackage{amsfonts}
\usepackage{amssymb}
\usepackage{graphicx}
\graphicspath{ {images/} }

\usepackage[final]{pdfpages}
\usepackage{tikz}
\usetikzlibrary{matrix} 
\usepackage{tikz}
\usetikzlibrary{arrows}
\usetikzlibrary{decorations, decorations.pathmorphing} 
\usetikzlibrary{decorations, decorations.pathreplacing}
\usetikzlibrary{backgrounds}
\usepackage{forest}
\usepackage{adjustbox}
\usetikzlibrary{calc,positioning,shadows.blur,decorations.pathreplacing} 
\usepackage{wrapfig}

\usepackage[toc,page]{appendix} 

\usepackage{multicol}
\usepackage{ geometry,  multirow,   amsthm, url, array}
\usepackage{tabularx}
\usepackage{longtable,booktabs}

\usepackage{colortbl}
\usepackage{xcolor}

\newlength\mylength
\newcolumntype{C}[1]{>{\centering\arraybackslash}p{#1}} 

\usepackage{chngcntr} 
\counterwithin{table}{section}
\counterwithin{figure}{section}

\numberwithin{equation}{section} 

\usepackage{biblatex} 
\usepackage{pgfplots}

\usepackage{url}
\usepackage{breakurl} 
\usepackage[breaklinks]{hyperref}

\usepackage{hyperref}
\hypersetup{
	colorlinks=true, 
	linktoc=all,     
	linkcolor=blue,
	filecolor=magenta,      
	urlcolor=cyan,
}
\usepackage{enumitem}
\makeatother
\theoremstyle{plain}

\theoremstyle{definition}

\theoremstyle{remark}

\title{{\bfseries {Sidechains: Decoupled Consensus Between Chains}}   \\ \medskip \large {Alberto Garoffolo and Robert Viglione} }

\author{Horizen - Zen Blockchain Foundation}
 \date{October 2018}
\usepackage{algorithm}
\usepackage{algorithmic}

\usepackage{etoolbox}
\makeatletter
\patchcmd{\@afterheading}%
    {\clubpenalty \@M}{\clubpenalties 3 \@M \@M 0}{}{}
\patchcmd{\@afterheading}%
    {\clubpenalty \@clubpenalty}{\clubpenalties 2 \@clubpenalty 0}{}{}
\makeatother

\usepackage[justification=centering]{caption}
\usepackage[margin=1cm]{caption}
\begin{document}

\maketitle
	
 \section*{ \centering }
 	\begin{center}
		   \textbf { Abstract}
		\end{center}
We propose a novel sidechain construction tailored to be compatible with the Horizen blockchain and designed for conducting secure and decentralized cross-chain transfers without requiring the mainchain nodes to track sidechains to verify them. The proposed scheme can also be adopted for other similar blockchain systems. We show that our cross-ledger transfer mechanism is secure under certain plausible assumptions. 
 
\section{Introduction}

The cryptocurrency phenomenon emerged in 2008 with Bitcoin \cite{11SNacam}, and since has gained a lot of traction among experts from various areas. Bitcoin was a first successful implementation of a decentralized payment system based on peer-to-peer networking. The key feature of Bitcoin - absence of a centralized control - is claimed to be a disruptive innovation that will help to build more robust, fair and transparent financial systems. 

Being fully open-sourced, Bitcoin inspired many other systems that leverage the same core principle of decentralization, but introduce a lot of additional advancements like smart contracts \cite{12EThnex}, private payments \cite{13ZCashweb,  14Moner}, decentralized governance \cite{15Dashwhit}, etc. Horizen \cite{16Horizenweb, 17ZenWhit} is an example of such a system. While retaining the basic blockchain functionality, it introduces a variety of new features such as private payments, decentralized governance, robust network of nodes (secure nodes), etc.

Since Horizen is based on the ZCash source code which in turn is based on Bitcoin, it retains all the constraints present in the original Bitcoin protocol like limited throughput,  increased latency, reduced ability to scale, etc. \cite{18ScalBL}. What's even more important is that such decentralized systems are very inert to change since there is no single entity to decide on updates. Even a tiny protocol alteration requires cumbersome process of agreement among community participants which makes introduction of new features very difficult. 

\begin{figure}[htbp]
	\centering
	\includegraphics[trim={3cm 9cm 8.5cm 2.5cm},clip,width=.65\linewidth] {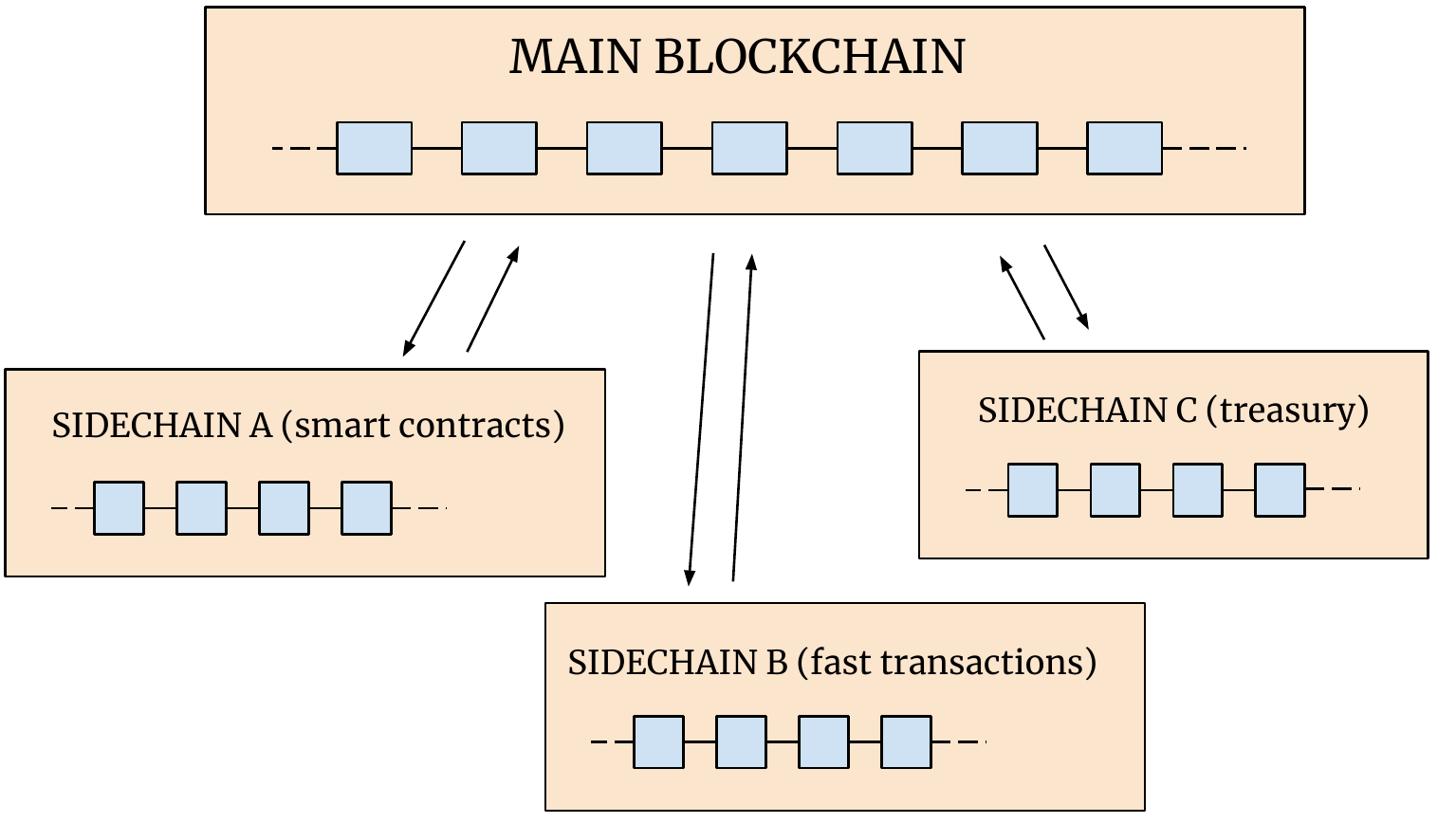}
	\caption{Basic concept of a sidechain. Users are allowed to transfer coins between the mainchain and a sidechian to get needed functionality.}
	\label{fig:1BAsConc}
\end{figure}

These problems forced researchers to look for possible solutions that would ameliorate such constraints. One of the most appealing, called sidechains, was presented by A. Back et. al. in 2014 \cite{1peg_sid1}. The sidechains approach enables to improve an existing blockchain system without actually changing the system itself. The basic idea is simple yet powerful: construct a parallel chain with whatever features are needed and provide a way to transfer value, or coins, between those chains (see Fig. \ref{fig:1BAsConc}). 

This way, for instance, a blockchain system like Bitcoin, which natively does not support smart contracts, may be extended with such functionality by leveraging the sidechain approach \cite{19Rootstk}. Actually, any number of sidechains can be deployed with different features and properties, while the main blockchain remains untouched.

The most crucial and controversial part in the sidechain design is the coin transfer mechanism (2-way peg or simply 2WP). Existing sidechain architectures provide different ways to implement 2-way pegs with different properties and tradeoffs.

In this paper we propose a novel sidechain construction tailored to be compatible with the Horizen blockchain (but actually it can also be adopted for other blockchain systems) and designed for conducting secure and decentralized cross-chain transfers without requiring the mainchain nodes to track a sidechain to verify them. The proposed sidechain construction requires modifications in the mainchain core consensus protocol to enable cross-chain transfers, but once implemented, it will allow deploying many sidechains in a reliable way.

The paper is structured in the following way: Section 1.1 provides an overview of the existing developments in this area. Section 2 provides more detailed information about the necessity of the sidechains in Horizen and basic requirements for its construction. Section 3 provides in-depth, technical details of the proposed construction.

\subsection{ Related work}

The concept of sidechains was first introduced by A.Back et al. in 2014 \cite{1peg_sid1}. This paper introduced a general notion of a 2-way peg and described two operational modes - synchronous and asynchronous - to implement interactions between pegged chains. The synchronous mode implies that both main and side chains are aware of each other and can verify transfer transactions directly, while the asynchronous mode relies on validators to process transfers.

Notable implementation of the general concept was presented in \cite{2wpeg_sid2,3hyb2peg_sid} and called Drivechains. It aims to deploy sidechains on top of the Bitcoin network. While forward transfers (from the mainchain to a sidechain) are processed by providing SPV proofs from the mainchain (like the synchronous mode in \cite{1peg_sid1}), backward transfers relied upon validators. The validators in the Drivechain proposal are the mainchain miners who need to track the sidechain and endorse transfers.

A. Kiayias and D. Zindros, in their presentation \cite{4Powsid}, proposed an implementation of the sidechain protocol for the proof-of-work blockchains by leveraging smart contracts. Another notable sidechain construction that relies on the smart contracts is called Plasma and was presented in \cite{6Plasma} by J. Poon and V. Buterin. On the contrary, our construction does not rely on the smart contract technology to provide a 2-way peg.

There have been many other attempts to construct cross-chain transfer mechanisms including Liquid project \cite{7Str_Fed}, Polkadot \cite{9Polkadot}, Interledger \cite{8int_pay}, Cosmos \cite{10Cosmosnet} and many others. They propose various solutions to implement 2-way pegs that are different from our construction and mostly targeted at private blockchains and federated transfers.
\section{Preliminaries}\label{sec:Pre}

The Horizen blockchain system has an extensive roadmap with a variety of new functionalities such as a treasury system for decentralized governance of resources \cite{20Tresur}, decentralized payment network for secure and super nodes, etc. Some of these functionalities require significant modifications of the core client, which, if implemented directly in the existing codebase, could have serious drawbacks:

\begin{enumerate}
	\item  Continuous growth of the monolithic application that exaggerates the overall architecture and complicates introduction of new features.
	
	\item  The existing C++ codebase is quite complex to change.
	
	\item  Integration of a new functionality into the core client comes with security risks.
	
	\item  A hard/soft fork is required to change the consensus protocol that cannot be done often due to a cumbersome process for community agreement.
	
	\item  Throughput/latency limitations of the core consensus protocol.
	
	\item  Limited flexibility to introduce new features (for instance, it's hard to substitute an existing consensus protocol with DAG-based one to allow fast transactions).
\end{enumerate}

For these reasons, it was decided to choose a sidechain approach for integration of new functionalities. It is one of the most appealing approaches because it allows overcoming at least the following constraints:

\begin{enumerate}
	\item  A sidechain implementation is completely decoupled from the mainchain, so no modifications of the core client are required (except the initial implementation of the sidechains support that is done only once). A sidechain may use whatever technology is suitable for implementing its distributed ledger, the mainchain is not required to know about it. 
	
	\item  Possible security impacts in the case of faulty implementation of the sidechain functionalities are bounded inside the sidechain and are limited only to the sidechain balance in the mainchain. 
	
	\item  Only one hard fork of the core consensus protocol is required (to enable sidechains). Then any new sidechain may be deployed seamlessly without changing the main network.
\end{enumerate}

Moreover, the sidechain approach generally allows much more flexibility. It can open a way for reliable integration of a variety of new features (for instance, simple or smart contracts, fast transactions etc).

\subsection{Requirements for the Horizen sidechains}

Considering the aforementioned constraints, we can outline the main requirements for the Horizen sidechains architecture:

\textbf{The mainchain should be agnostic to the sidechains.} The mainchain nodes should not be required to track a sidechain. Only the cross-chain transfers are known and verified by the mainchain, but to verify them it should be enough to track only the mainchain.

\textbf{Unified cross-chain transfers.} All sidechains should employ the same unified cross-chain transfer (CCT) protocol that is known by the mainchain. Even though the participants of the CCT protocol will be different for different sidechains, the procedure should be specified by the mainchain consensus protocol. Such unification will allow the system to deploy many sidechains without the need to modify the mainchain consensus protocol for each of them separately.

\textbf{Flexibility to choose sidechain consensus protocol.} Any sidechain consensus protocol may be different from the one used by the mainchain. Generally, the mainchain should not rely on any assumptions regarding the sidechain consensus protocol or its structure. Consensus about cross-chain transfers should be decoupled from the specific sidechain consensus protocol.

\textbf{Parallel sidechains.} The Horizen sidechains architecture shouldn't restrict the number of simultaneously deployed sidechains. All deployed sidechains should work in parallel completely independently. Cross-chain transfers should also be completely independent for different sidechains.

\textbf{Fault tolerance.} The mainchain should be fault tolerant to any sidechain failure (including malicious behaviour). The risk should be limited only to the sidechain balance. It should be impossible to withdraw more coins from the sidechain than were previously received, even in the case of total sidechain corruption.

\textbf{Minor changes in the core.} The required modifications in the Horizen core should not be too pervasive.

The \textbf{Sidechains SDK} is planned to be developed such that it will provide out-of-the box implementation of the template sidechain with pure consensus protocol. The interfaces should be designed in such a way that the main components could be easily replaced.
\section{The sidechain construction}\label{sec:SidCon}

This section provides details of the proposed sidechain construction. We start by defining the abstract model of a sidechain, and then specify the details of the constituents of the proposed structure.

\subsection{ General overview}\label{sec:GenOver}

Analyzing existing attempts to construct sidechains \cite{1peg_sid1,2wpeg_sid2,3hyb2peg_sid,4Powsid}, we may outline two major components of any sidechain structure:

\begin{enumerate}
	\item  \textbf{Cross-chain transfer protocol} \textbf{- CCT} (2-way peg).
	\item  \textbf{Sidechain consensus protocol - SCP.}
\end{enumerate}

Defining these two components becomes crucial in specifying an exact sidechain construction. Depending on the implementation, they can be incorporated into the mainchain or completely independent of the mainchain logic.

In our construction, we decouple these components. The \textbf{CCT} protocol is to be unified and fixed by the mainchain logic, so that all sidechains will use the same \textbf{CCT} protocol. The \textbf{SCP} protocol will be completely decoupled from the \textbf{CCT} and mainchain logic in general, so that a sidechain developer is free to choose the \textbf{SCP} protocol depending on needs and preferences.

Even though the SCP protocol may vary for different sidechains, we are going to provide a reference construction.

\textbf{Cross-chain transfer protocol. }The CCT protocol defines the structure of the 2-way peg, so it consists of two sub-protocols. The first defines a \textit{forward transfer} procedure and the other a - \textit{backward transfer }procedure. A \textit{forward transfer }transmits coins from the mainchain (\textbf{MC}) to the sidechain (\textbf{SC}). A \textit{backward transfer} transmits coins back from SC to MC.

The CCT protocol is the most important part of the sidechain construction, since it defines the overall structure of the communications between MC and SC. The backward transfers are especially important, because the MC usually does not keep track of the sidechain so is not able to directly verify the validity of a SC to MC transfer.

In our construction, the forward transfers will be implemented by means of a special MC transaction that burns coins in the MC and provides metadata that allows a user to claim a corresponding amount of coins in the SC. Since the transfer is initiated on the MC and does not require any additional info from the SC, it can be easily validated by both MC and SC. The structure of the forward transfers is relatively straightforward in this case.

A more complex procedure is required for doing backward transfers. They are initiated in the SC, and then propagated to the mainchain while burning the corresponding amount of coins on the SC and re-creating them on the MC. Since the MC does not follow the SC, it cannot verify directly the validity of such backward transfers. To keep the ability of the MC to validate backward transfers, we employ a set of special actors called \textbf{certifiers} who manage the certification of such transfers. The certifiers register themselves in the MC by staking particular amount of coins. Their main purpose is to track the sidechain, collect all backward transfers into certificates, sign certificates and propagate them to the mainchain. Since the list of certifiers is known to the mainchain, the certificates can be easily verified.

\textbf{Sidechain consensus protocol.} Despite the fact that the SCP can be implemented in different ways, we are going to provide a reference structure. It is based on the scheme proposed in Ouroboros \cite{21POsprovsec} and modified to provide binding with the mainchain blockchain. It also suggests a randomness generation algorithm different from one proposed in Ouroboros.

In the following sections, we provide a detailed description of the mentioned protocols. We will start with the SCP description and then build the CCT protocol upon it.

\subsection{ Sidechain consensus protocol}\label{sec:SidCon_pr}

\begin{figure}[htbp]
	\centering
	\includegraphics[trim={1.5cm 10.5cm 2.5cm 4.5cm}, clip,width=.8\columnwidth]{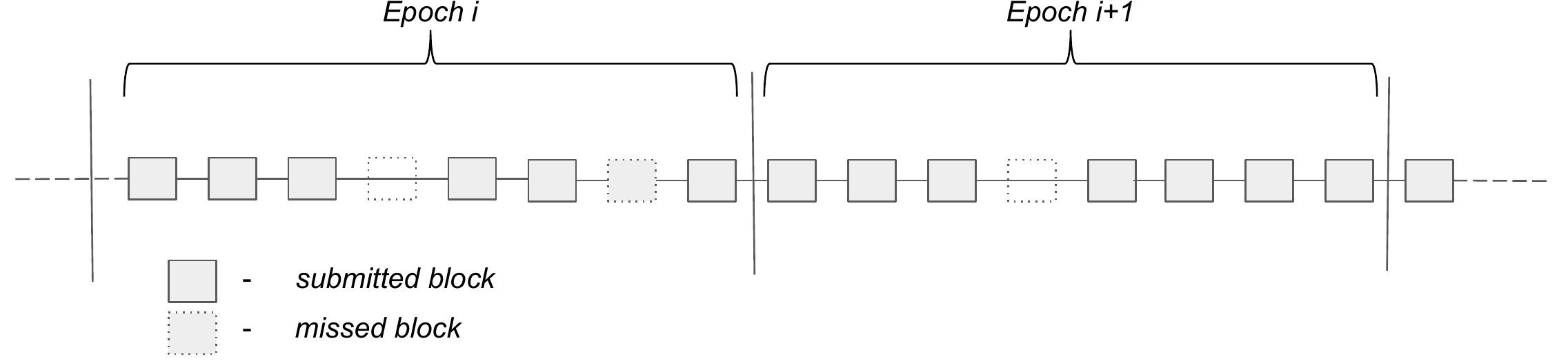}
	\caption{A general scheme of an epoch. Note that even though there is an assigned slot leader for each slot, the leader may skip block generation and the slot remains empty.}
	\label{fig:2allst}
\end{figure}

For the sidechain consensus protocol, we adopted a modified version of the Ouroboros proof-of-stake protocol \cite{21POsprovsec}. Its main idea is that time is divided into epochs with a predefined number of slots. Each slot is assigned to a slot leader who is authorized to generate a block during this slot. Slot leaders of a particular epoch are chosen randomly before the epoch begins from the set of all sidechain stakeholders (Fig. \ref{fig:2allst}). The protocol operates in a synchronous environment where each slot takes a specific amount of time (e.g. 20 seconds).

\textbf{Epoch}. An epoch is a sequence of the $k$ successive slots at particular positions ${Ep}_i=({sl}^0_i,{sl}^1_i,...,{sl}^{k-1}_i)$, where $k$ is the predefined length of the epoch and $i$ is the sequence number of the epoch.

\textbf{Slot.} A slot is a specific period in time during which a slot leader is authorized to issue a block. Each slot has a corresponding slot leader who is chosen randomly before the epoch begins. A slot leader may skip generating a block, and in this case the following block will refer to the latest generated block.

\textbf{Slot Leader.} A slot leader of the slot ${sl}^j_i$ is a stakeholder who was authorized by the Slot Leader Selection Procedure to forge a block at slot ${sl}^j_i$.

\textbf{Slot Leader Selection Procedure.} A slot leader selection procedure $Select({SD}_{{Ep}_i},\ rand)$ is a procedure that selects all slot leaders of the epoch ${Ep}_i$ according to the fixed stake distribution ${SD}_{{Ep}_i}$ and some random value $rand$. The stake distribution ${SD}_{{Ep}_i}$ is fixed before the epoch ${Ep}_i$ begins. The randomness $rand$ is revealed only after the stake distribution is fixed.

In our construction, we modified the protocol to introduce binding with the mainchain. It means that a sidechain block may contain reference(s) to the mainchain block(s) so that the history of the mainchain blocks is stored in the sidechain. By reference, we understand the hash of the MC block or, in the case that the MC block contains transactions related to this sidechain, the whole block header together with SC transactions and their Merkle path.

\begin{figure}[htbp]
	\centering
	\includegraphics[trim={1.5cm 10.5cm 2.5cm 6.5cm}, clip,width=.86\columnwidth]{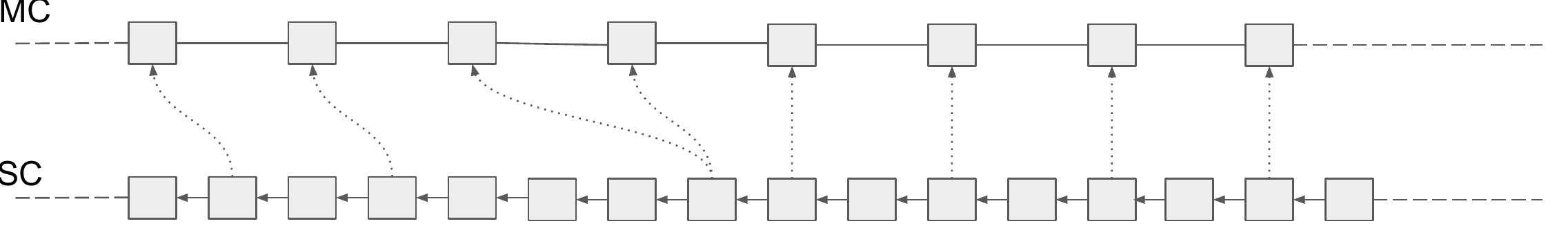}
	\caption{An example of the sidechain binding to the mainchain blocks}
	\label{fig:3genref}
\end{figure}

The sidechain block forgers are obliged to keep mainchain references consistent and ordered when included into the  SC blocks. A sidechain block ${SB}_j$ can contain a reference to the mainchain block $B_i$ if and only if (1) the block $B_i$ is a valid mainchain block, and (2) the references to all previous mainchain blocks $B_k,\ k\in \{\eta ,\eta +1,...,i-1\}$ are already included into the sidechain blocks (including the current one as it may contain more than one reference), where $\eta $ is the genesis reference (Fig. \ref{fig:3genref}).

Note that it is not mandatory for the block forgers to include mainchain references, but we assume that honest block forgers will do this to support coin transferring between SC and MC. It is also possible to construct an incentive mechanism for the block forgers who include references. For instance, the users who initiate forward/backward transfers may pay a small fee from each transaction. The incentive mechanism is beyond the scope of the current research as we provide only an example of sidechain consensus protocol.

There are two main reasons to introduce SC to MC binding:

\begin{enumerate}
	\item  \textbf{Deterministic synchronization between MC and SC. }When a sidechain block ${SB}_i$ refers to the mainchain block $B_j$, it explicitly acknowledges all transactions from the block $B_j$ and from all previous blocks. It means that if $B_j$ contains any transactions related to this sidechain (for instance, ${MC \rightarrow SC}$ coin transfer), such transactions may be immediately transferred to the sidechain. \textbf{}
	
	\item \textbf{ Finality of the mainchain is not guaranteed. }\textit{In the following considerations we assume that the mainchain is the Horizen blockchain which uses modified Nakamoto consensus with a block delay penalty mechanism.} It is known that Nakamoto consensus does not provide finality on the chain of blocks \cite{22FormalConsen}. It means that there is always non-zero probability that some sub-chain of blocks will be reverted and substituted by another chain with more cumulative work. Such behaviour is normally handled by the mainchain but may be disastrous for the sidechain because ${MC \rightarrow SC}$ transactions that are already confirmed in the sidechain may be reverted in the mainchain. The binding eliminates such a situation, because in the case of a fork in MC, the SC blocks that refer to forked blocks in MC would also be reverted.
\end{enumerate}

\begin{figure}[htbp]
	\centering
	\includegraphics[trim={6cm 10.5cm 6cm 6.5cm}, clip,width=0.8\columnwidth]{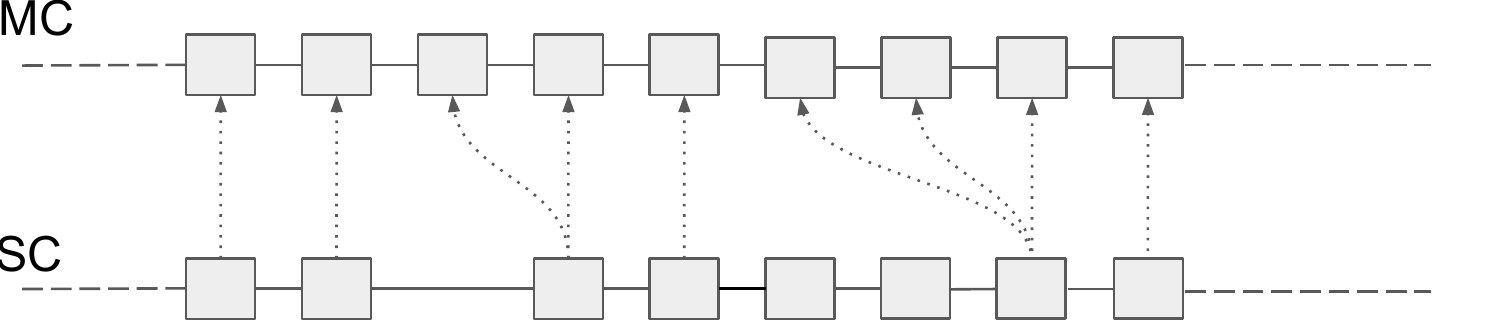}
	\caption{Example of missing mainchain block references}
	\label{fig:4Exmiss}
\end{figure}

\textbf{Full referencing. }Full referencing implies that the sidechain blocks contain the full chain of the mainchain block references. Even if some block forger missed an opportunity to include a reference to the newly generated MC block, some of the following block forgers will include the missed mainchain reference (see Fig. \ref{fig:4Exmiss}).

The reference itself is a hash of a MC block or complete block header of the corresponding mainchain block in the case it contains cross-chain transactions. This way it becomes possible to do lite verification of the mainchain blocks that allows the introduction of additional features in the sidechain, such as lite verification of the cross-chain transactions. For example, when a transfer transaction ${tx_{MC \rightarrow SC}}$ happens in the mainchain block $B_j$, a sidechain slot leader creates the corresponding sidechain block, where he includes, beside the header of the $B_j$, the transaction $tx_{MC\rightarrow SC}$ and the Merkle path of this transaction. In this case, such a transfer can be verified by any sidechain node without necessity to parse the mainchain.

\begin{figure}[htbp]
	\centering
	\includegraphics[trim={3cm 11.2cm 3cm 1.5cm}, clip,width=.86\columnwidth]{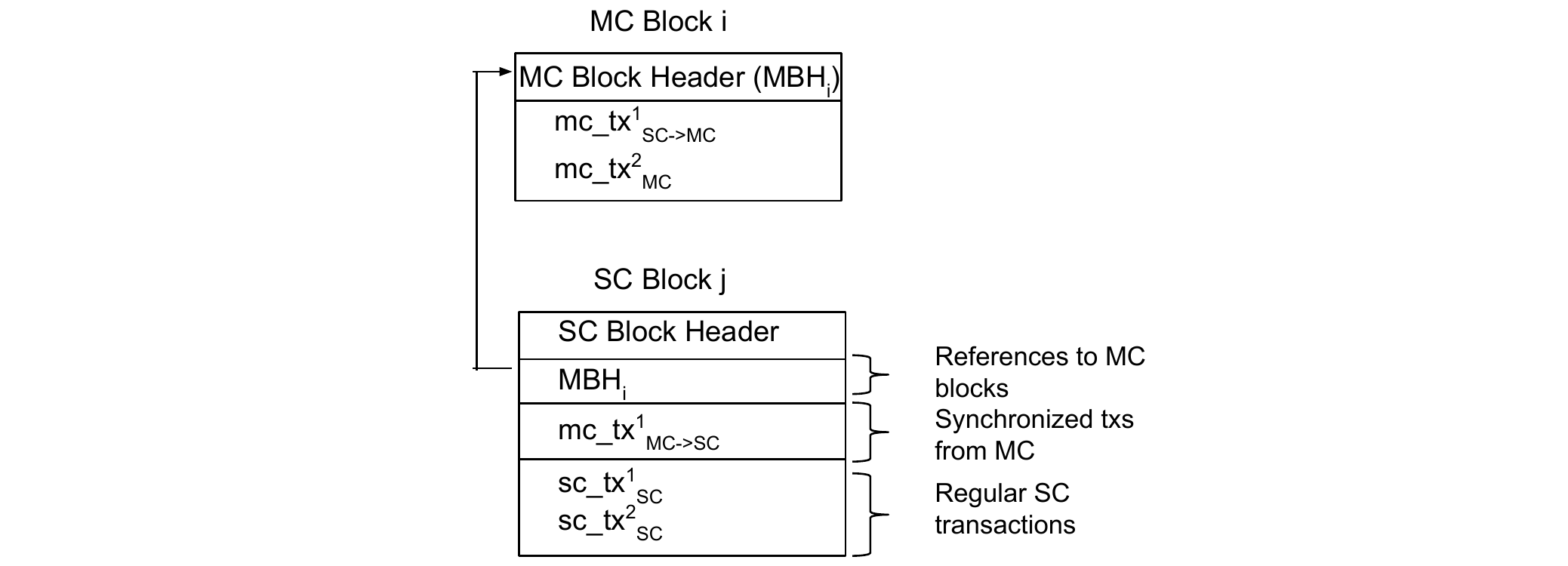}
	\caption{Example of txs synchronization between the mainchain and a sidechain}
	\label{fig:5Extxs}
\end{figure}

\textbf{Deterministic synchronization.} By deterministic synchronization we imply the rule that a sidechain slot leader is obliged to include into a generated sidechain block all transactions related to this sidechain that appeared in the mainchain blocks to which he refers (Fig. \ref{fig:5Extxs}).

In this way cross-chain transactions from MC are immediately synced to the sidechain. If a slot leader breaks the rule and does not synchronize transactions from the mainchain, such a block will be deemed invalid and should be forked by the successive blocks.

Together with a synchronized transaction, its Merkle path should be provided to the root that is stored in the referred mainchain block header.

\textbf{Randomness.} The crucial component of the Ouroboros proof-of-stake consensus protocol is the truly random, unbiased source of randomness. Security is guaranteed only if randomness is present. In the original Ouroboros the randomness is obtained through the special coin-tossing protocol based on verifiable secret sharing. This protocol is run by the slot leaders.

In our sidechain construction, we leverage the presence of a proof-of-work mainchain and derive randomness from the proof-of-work solutions. The exact protocol and its security will be described in further sections.

\textbf{Security.} The standard procedure of proving blockchain consensus protocol security requires demonstrating ability of the protocol to satisfy two fundamental properties of a distributed ledger: \textbf{liveness} and \textbf{persistence }\cite{23GKL}. Liveness ensures that the transactions broadcast by the honest parties will be eventually included into the ledger, and persistence ensures that once a transaction is confirmed by one honest node, it will be also confirmed by all other honest nodes (so that eventually it becomes final and immutable). Such properties are usually proven under certain assumptions such as honest majority among protocol participants. We refer interested readers to the original Ouroboros paper \cite{21POsprovsec} for the exhaustive list of assumptions with their strict definitions.

Since the proposed consensus protocol also incorporates binding with the mainchain blocks, it implies the additional assumption of the honest hashing power majority in the referred chain.

We suppose that under these assumptions the proposed protocol derives the security guarantees provided by the original Ouroboros and Nakamoto consensus protocols.

We want to emphasize that different sidechains may adopt different consensus protocols that better suit specific use cases (e.g. fast coins transferring support). A sidechain's consensus protocol (including the one described in this section) is not the main focus of this research and needs further rigorous security analysis.

\subsection{Cross-chain transfer protocol}\label{sec:CrtrPr}

This section describes the cross-chain transfer protocol which is the most crucial part of the sidechain construction. The CCT protocol basically consists of two components:

\begin{enumerate}
	\item  \textbf{Forward Transfers Protocol.} Defines transfers from the mainchain to the sidechain.
	
	\item  \textbf{Backward Transfers Protocol. }Defines transfers from the sidechain to the mainchain. 
\end{enumerate}

We also distinguish the \textbf{Certification Protocol} in a separate section. Logically, the Certification Protocol is a part of the Backward Transfers Protocol, since it is designed to verify transfers from a sidechain to the mainchain, but for the sake of clearer architecture and to facilitate the security analysis we describe it separately.

\subsubsection{Forward Transfers Protocol}\label{sec:FortrPr}

The design of the forward transactions is straightforward and basically the same as in many of the existing sidechain architectures \cite{1peg_sid1, 2wpeg_sid2, 3hyb2peg_sid}. 

\begin{figure}[htbp]
	\centering
	\includegraphics[trim={1cm 7.7cm 0.5cm 4.2cm}, clip,width=0.86\columnwidth]{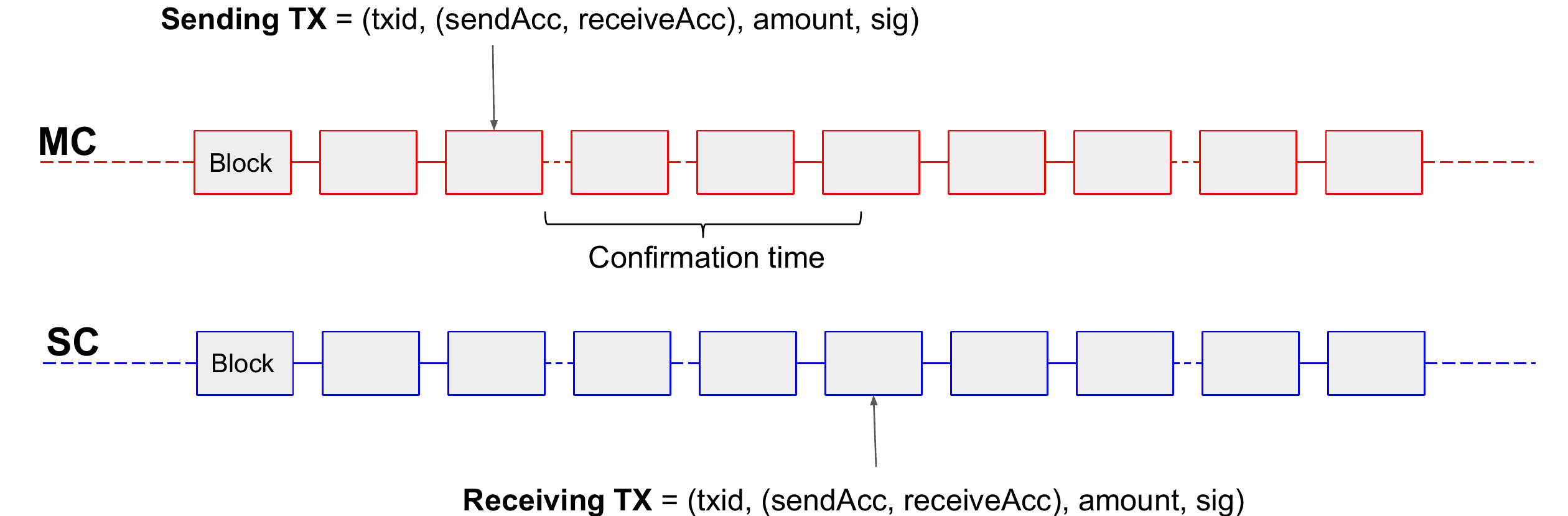}
	\caption{Standard approach for the MC to SC transfer}
	\label{fig:6StanAp}
\end{figure}

In general, it looks as follows: MC to SC transfer (Fig. \ref{fig:6StanAp}) is represented by a pair of transactions, the one on the MC (Sending TX) and another one on the SC (Receiving TX). The purpose of the Sending TX is to lock/burn tokens on the MC. The purpose of the Receiving TX is to unlock/create corresponding amount of tokens on the sidechain. The Receiving TX is valid only in the case of the confirmed Sending TX (basically Sending TX is a proof of locked coins).

This transfer mechanism can be easily modified to leverage the advantage of the presented sidechain consensus protocol (section \ref{sec:SidCon_pr}). Instead of relying upon a user to create Receiving TX in the sidechain, it can be automatically included by the sidechain block forger who refers to the mainchain block with Sending TX. 

In what follows, we define the forward transfer more formally assuming that the sidechain consensus protocol is the one described in section \ref{sec:SidCon_pr}.

\textbf{Definition (Sending Transaction).} Sending transaction is a specially structured transaction that initiates a transfer of coins from the original blockchain \textbf{\textit{A }}(the mainchain) to the destination blockchain \textbf{\textit{B }}(the sidechain). The basic structure of the Sending TX is defined as follows:

\[Sending\ TX\ =\ \{ledgerid,\ txid,\ (sendAcc,\ receiveAcc),\ amount,\ sig\},\] 

where 

\textbf{\textit{ledgerid}} is a unique identifier of the previously deployed sidechain to which the transactions are targeted;

\textbf{\textit{txid}} is a unique transaction identifier;

\textbf{\textit{sendAcc}} is an address on the original blockchain \textbf{\textit{A}} for which $balance(sendAcc)\ge amount$

\textbf{\textit{receiveAcc}} is  an address on the destination blockchain \textbf{\textit{B}};

\textbf{\textit{amount}} is the number of coins to transfer;

\textbf{\textit{sig}} is  a signature that corresponds to the \textit{sendAcc}.

The Sending TX burns \textit{amount} of coins on the original blockchain \textbf{\textit{A }}where it belongs, and reduces the \textit{sendAcc} balance by this amount.

\begin{figure}[htbp]
	\centering
	\includegraphics[trim={1cm 4cm 0.5cm 4cm}, clip,width=.86\columnwidth]{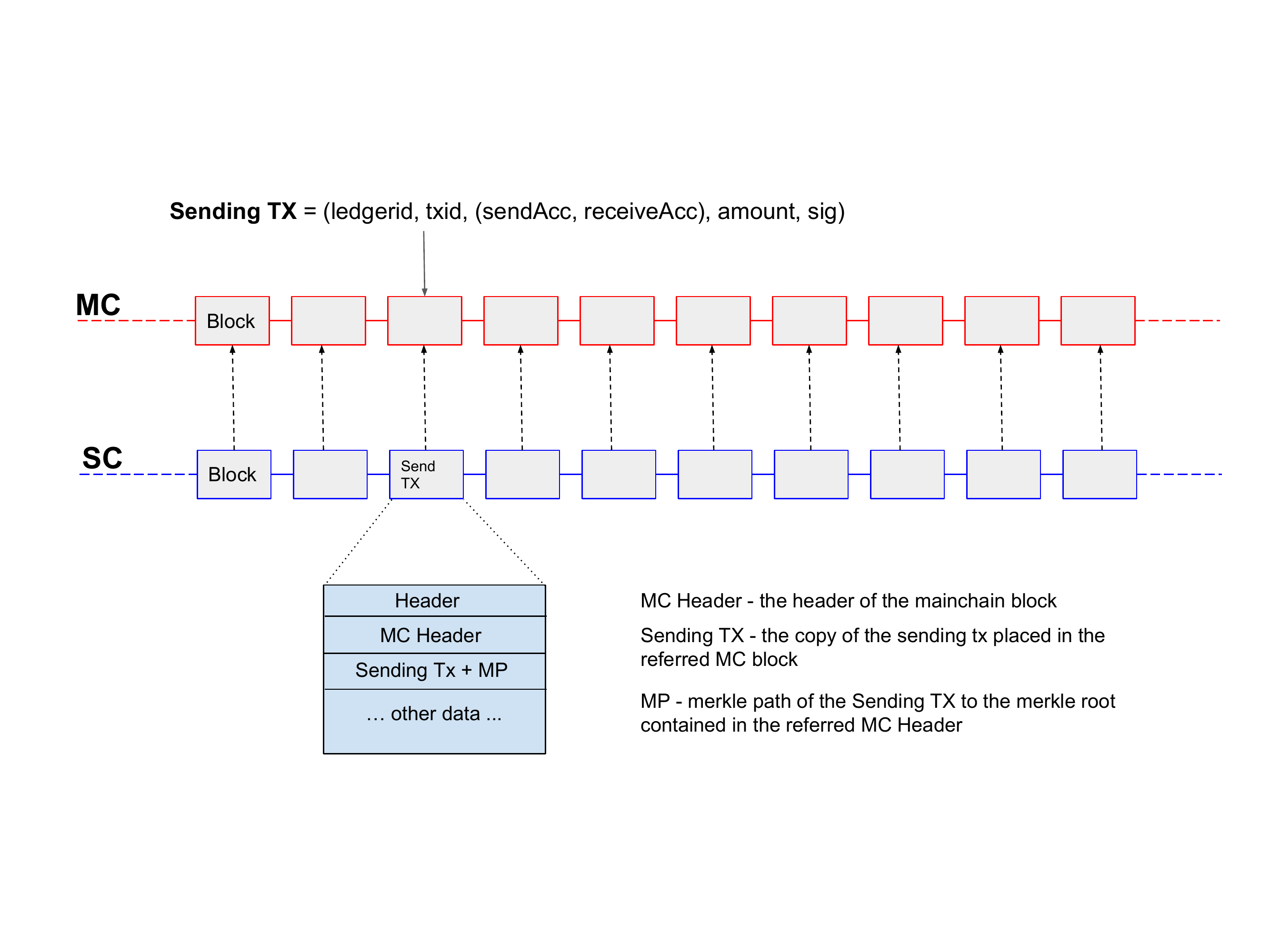}
	\caption{ Forward transaction syncing from mainchain to sidechain}
	\label{fig:7Forwtrx}
\end{figure}

A special receiving transaction is not needed in our case, because information about the transfer that has been submitted with the Sending TX is included into the sidechain by a block forger who refers to the block with the Sending TX (see Fig. \ref{fig:7Forwtrx}). To guarantee the possibility to perform lite verification of the sending transaction in the sidechain block (without the necessity to track the mainchain), the full Merkle path of the sending transaction is also included. Since the Merkle root is present in the mainchain block header (which is also included into the SC block), one could easily verify the sending transaction in the sidechain.

After inclusion into the sidechain block, the corresponding \textit{amount} of coins is created and the \textit{receiveAcc} balance is increased by this amount.

Note that the provided transaction structure is not exhaustive. It only defines the general information that should be included in the transaction. The specific format may vary depending on the transactions model of the blockchain systems. For instance, for the Horizen system - that utilizes the Bitcoin transaction model - the Sending TX can be implemented by introducing a special opcode like OP\_SENDSIDECHAIN or by introducing a completely new type of transaction. It is up to a developer to decide the most suitable path.

\textbf{ Security.} To analyze the security of the described scheme, we will introduce a property called \textbf{forward transfer safety} that requires that no coins can be created on the sidechain unless corresponding amount of coins have been burnt with the Sending TX in the mainchain. 

It is easy to see that if the consensus protocol with tight binding is used (as described in Section \ref{sec:SidCon_pr}), the forward transfers protocol completely satisfies the safety property. If the Sending TX is reverted in the mainchain, the corresponding sidechain block becomes invalid and is also reverted. Moreover, only the creator of the Sending TX may specify the receiver address. To falsify a transfer, one would need to forge the signature of the \textit{sendAcc }which is assumed to be infeasible.

\subsubsection{Backward Transfer Protocol} \label{sec:BActrPr}

The backward transfer protocol is the most complex part of any sidechain construction. This is the component that differentiates this research from many other existing constructions. The complexity stems from the assumption that the mainchain knows nothing about a specific sidechain, so it is impossible to directly verify the transfer that has been initiated in the sidechain. 

There are different approaches to implement backward transfers. For instance, A.Back et al. in their construction \cite{1peg_sid1} rely on the federation of authorized entities to verify transfers, the Drivechain approach \cite{2wpeg_sid2, 3hyb2peg_sid} involves mainchain miners to vote (and thus verify) the correctness of the backward transfers.

In our construction, we rely on the idea of a batch verification of the backward transfers (in some sense the same as in \cite{3hyb2peg_sid}), but the verification process is different from the existing constructions. Instead of relying upon the miners (block forgers), we introduce independent entities called \textbf{certifiers,} who are responsible for verification of correctness of the backward transfers. The certifiers are registered by locking of their stake for the period of operation.

The remainder of this section provides details on how backward transfers are organized, verified and transferred to the mainchain. Section \ref{sec:Cert} will provide details about the certifiers selection protocol.

\begin{figure}[htbp]
	\centering
	\includegraphics[trim={1cm 9cm 2.5cm 5.5cm}, clip,width=.86\columnwidth]{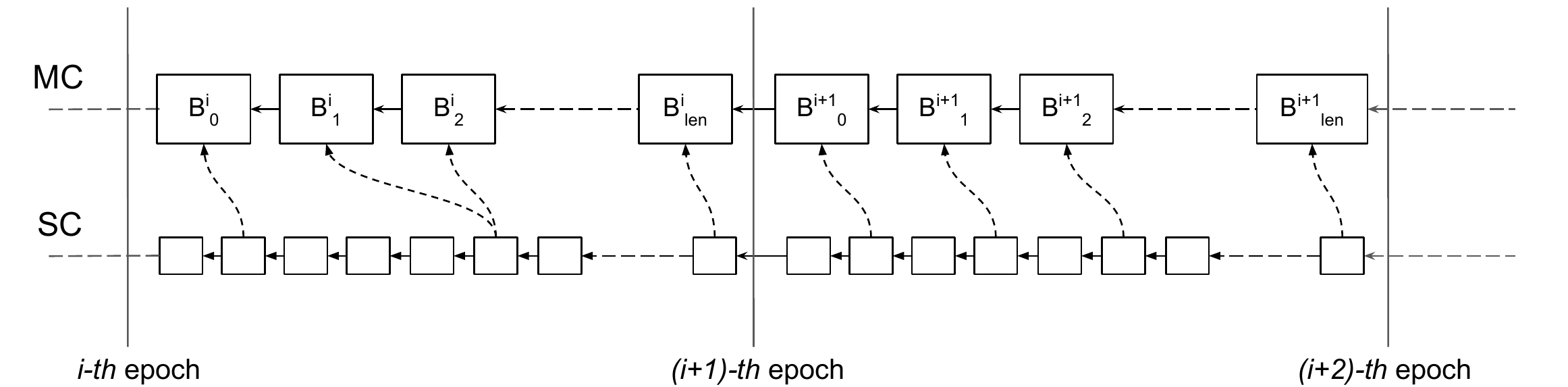}
	\caption{Withdrawal epochs (a)}
	\label{fig:8aWithdrEp}
\end{figure}

The whole lifetime of the MC blockchain (and correspondingly of the SC) is divided into withdrawal epochs (not to be confused with SC consensus epochs - they are different) that have a determined number of blocks $epoch\_len$ (e.g. $epoch\_len$\textit{ = 720}). Let's define a block $B_i$ that belongs to a specific epoch \textit{ep\_id }as $B^{ep\_id}_j$, where $j\in 0\ ..\ epoch\_len$. In the sidechain, the withdrawal epoch (and any other sub-stages) is defined by the MC blocks' references, so that the epoch $ep\_id$ is started from the SC block that refers to $B^{ep\_id}_0$ and ends with the block that refers to $B^{ep\_id}_{epoch\_len-1}$ (Fig. \ref{fig:8aWithdrEp}).

\begin{figure}[htbp]
	\centering
	\includegraphics[trim={1cm 9cm 2cm 5.5cm}, clip,width=.86\columnwidth]{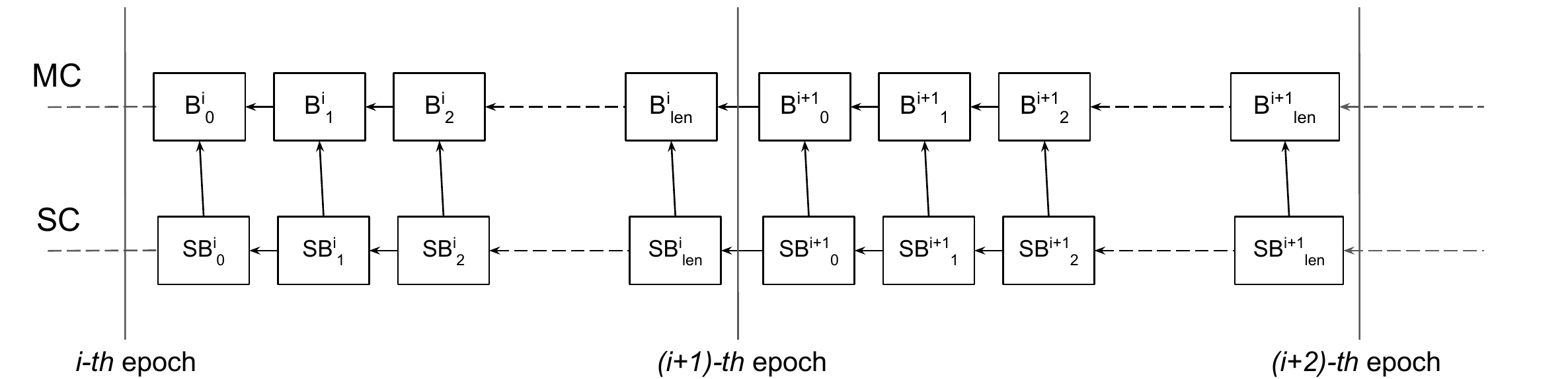}
	\caption{Withdrawal epochs (b)}
	\label{fig:8bWithdrEp}
\end{figure}

In the remainder of the document, without loss of generality and to simplify description, we will assume that the block generation rate is the same for MC and SC, and each SC block references exactly one MC block. We will use $B^{ep\_id}_j$ notation for mainchain blocks and ${SB}^{ep\_id}_j$ for the corresponding sidechain block that refers to $B^{ep\_id}_j$ (Fig. \ref{fig:8bWithdrEp}).

The main goal of having epochs is to provide an efficient mechanism to send authenticated data from a sidechain to the mainchain. Since the mainchain itself is not able to verify the authenticity and validity of the data, special entities called \textbf{certifiers} are empowered to certify the data transmitted. Certification of each message separately seems inefficient with the existing infrastructure, so the bulk certification approach is taken which implies that the data are first accumulated during the epoch, grouped into the structures called Cross-Chain Certificates, validated and signed by certifiers and only then transmitted to the mainchain. One such cycle of signing takes one withdrawal epoch.

\textbf{Definition (Cross-Chain Certificate).} \textit{Cross-Chain Certificate (CCCert) is a generic container that is used to transmit data from a sidechain to the mainchain. It is defined by the mainchain rules and consists of three major parts: Backward Transfers List, Certifiers Withdrawals List and Fraud Reports List.} \textit{To be accepted by the mainchain a CCCert should be signed by the majority of the authorized Certifier Group. More specifically, CCCert contains the following information:}

\begin{enumerate}
	\item \textit{ sidechain identifier;}
	\item \textit{ epoch number;}
	\item \textit{ number of the certificate in a particular epoch.}
	\item \textit{ Backward Transfer List}
	\item \textit{ Certifiers Withdrawals List}
	\item \textit{ Fraud Reports List}
\end{enumerate}

\begin{figure}[htbp]
	\centering
	\includegraphics[trim={5cm 12.3cm 7cm 2.5cm}, clip,width=0.6\columnwidth]{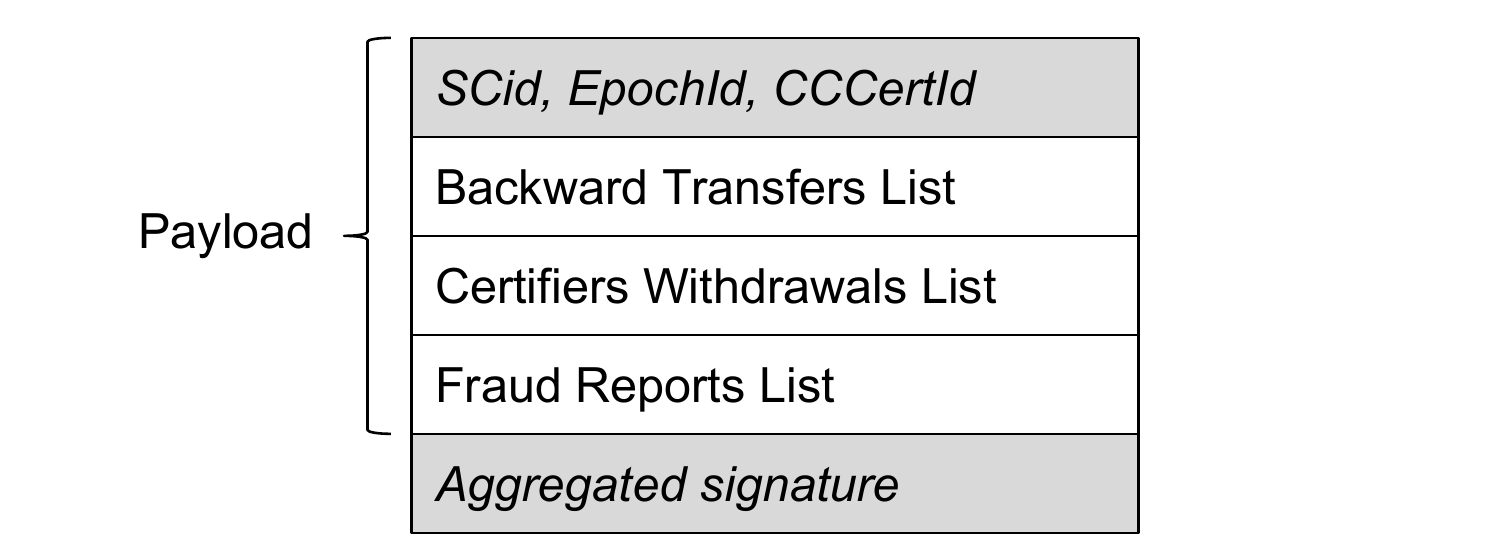}
	\caption{Cross-Chain Certificate basic structure}
	\label{fig:9CrCHain}
\end{figure}

The certifiers are registered in the mainchain, so the aggregated signature can be verified by the mainchain. Section \ref{sec:Cert}  will provide details about the certifiers' registration and selection. Note that each withdrawal epoch may generate up to  $k$  cross-chain certificates, where  $k$  depends on the number of certifiers.

Before proceeding with the basic backward transfer flow, we will define the notions of the Backward Transfer and Backward Transfers List which is a part of CCCert.

\textbf{Definition (Backward Transfer). }\textit{The Backward Transfer (BT) is an abstract structure that specifies details of a single transfer of coins from a sidechain to the mainchain. The structure is defined by the mainchain rules. In general, it must specify at least an amount of transferred coins and information about the receiver: BT(amount, receiver).}

More specifically, a backward transfer contains information about the receiver of coins in the mainchain. For instance, in the case of the Horizen blockchain system, it may contain an output script that defines ownership of the transferred coins in the Horizen mainchain.

The backward transfers for the given epoch are derived from the sidechain. The most common case for initiating a backward transfer is a withdrawal transaction created by a sidechain user who wants to transfer his coins to the mainchain. However, it is not limited to only users' withdrawal transactions. For instance, it may be a payment from the treasury fund, or any other payment defined by the sidechain rules.

\textbf{Definition (Backward Transfers List). }\textit{The Backward Transfers List (BTList) is a structure defined by the mainchain rules that contains a set of backward transfers for the total amount }$\sum^{|BT|}_{i=0}{}amount_{BT_i}\le MAX\_CERT\_AMOUNT$\textit{, where MAX\_CERT\_AMOUNT is the maximum allowed transfer amount for a single CCCert. The set of such lists is defined deterministically from the ordered set of backward transfers.}

\begin{figure}[htbp]
	\centering
	\includegraphics[trim={1cm 5cm 1cm 3.5cm}, clip,width=.86\columnwidth]{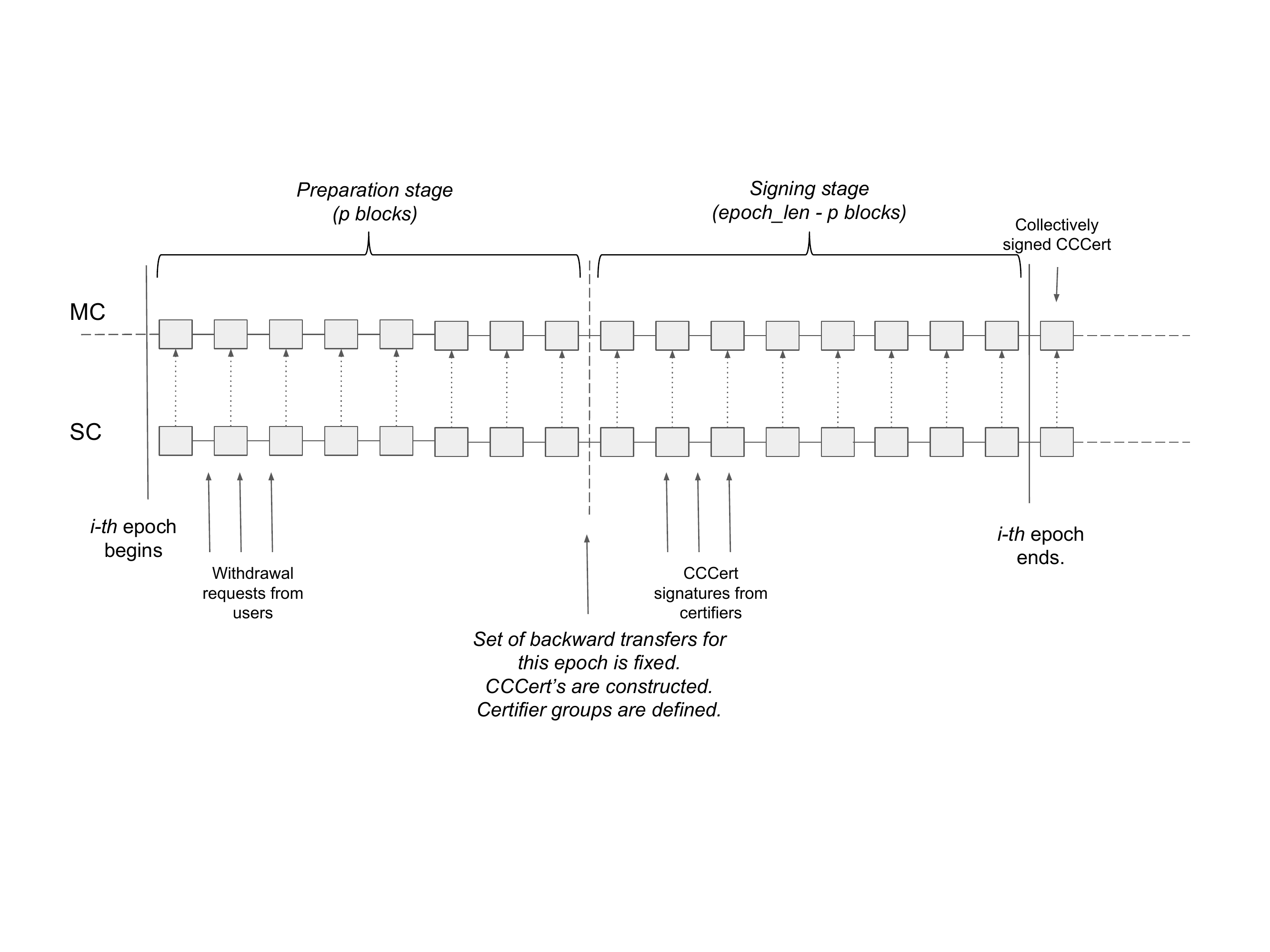}
	\centering
	\caption{Basic withdrawal flow}
	\label{fig:10BasWif}
\end{figure}

Each withdrawal epoch comprises two stages: \textbf{preparation stage} and \textbf{signing stage}. During the preparation stage, users are able to submit withdrawal requests that will be processed in that epoch. If a user submits withdrawal requests after the preparation stage, it will be processed only in the following epoch. The basic withdrawal flow is shown in Fig. \ref{fig:10BasWif}.

Certification of backward transfers for the epoch \textit{i }starts immediately after the block ${SB}^{ep\_id}_p$. At this point, the set of current backward transfers is fixed, the list of cross-chain certificates is constructed and each CCCert is assigned with a certifier group that should sign it.

The list of all backward transfers during the epoch is collected and sorted in chronological order. Then all backwards transfers are grouped into backward transfers lists according the Algorithm \ref{algo1group}.
\begin{algorithm}[!th]
	\caption{Grouping of backward transfers into lists}
	\label{algo1group}
	\begin{algorithmic}[1]
	\STATE   $BTList_{0}$  takes the first  $k_{0}$  backward transfers from the list until $\sum^{k_0+1}_{i=1}{}$ $amount_{BT_i}$   $>MAX\_CERT\_AMOUNT$ or until the list of backward transfers is exhausted. 
	
	\STATE    $BTList_{1}$  takes the following  $k_{1}$  backward transfers from the list until $\sum^{k_0+k_1+1}_{i=k_0+1}{}$ $amount_{BT_i}>$ $MAX\_CERT\_AMOUNT$ or until the list of backward transfers is exhausted. 
	
	\STATE   The process is repeated until the list of backward transfers is exhausted.
	\end{algorithmic}
\end{algorithm}

To limit the impact on the MC block in terms of the number of transactions related to the backward transfers, we introduce a minimum amount to be transferred 

\begin{center}{$amount_{BT_i}\ge \ MIN\_TRANSFER\_AMOUNT$.}\end{center}

In such a way, the maximum number of transfers per certificate would be 

\begin{center}{$m = \frac{MAX\_CERTIFICATE\_AMOUNT}{MIN\_TRANSFER\_AMOUNT}.$}\end{center}

Simply speaking, all backward transfers are grouped into BTList's in chronological order so that each  $BTList_{i}$  total amount does not exceed the $MAX\_CERT\_AMOUNT$ threshold. Then the  $BTList_{i}$  is included into the corresponding  $CCCert_{i}$ . The number of certificates in a particular withdrawal epoch depends on the number of backward transfers and the number of certifiers. If all backward transfers can be processed within a single certificate, then only one certificate will be created in the epoch. But only up to $\left[\frac{n}{N}\right]$ certificates can be issued, where $n$ is the number of available certifiers in the epoch and $N$ is the size of the certifier group.

Note that a particular withdrawal epoch may not have enough CCCert's to process all BTList's. In this case unprocessed backward transfers will be transmitted to the following withdrawal epoch.

All operations are done in a deterministic way without any additional on-chain transactions, so that each certifier exactly knows what he should sign, and each verifier later on knows how the certificate exactly should be constructed and who should sign it. During the signing stage, certifiers submit their signatures to the sidechain. After the epoch is finished, all signatures are aggregated, and certificates with aggregated signatures are pushed to the mainchain. 

The $MAX\_CERT\_AMOUNT$ value is equal to $\frac{\sum^N_{i=1}{}deposit(C_i)}{2}$, where $N$ is the size of the group which signs the certificate and $deposit(C_i)$ is the deposit amount of the certifier $C_i$. Simply speaking, the withdrawal amount per certificate is at most half of the total deposit amount of the certifiers eligible to sign this certificate. Since all certifiers deposits are equal (as will be discussed in further sections) and the size of the certifiers group is also constant, the $MAX\_CERT\_AMOUNT$ will have a constant value for all cross-chain certificates ever issued in the sidechain.

The above rule also implies that the maximum withdrawal limit per epoch is equal to half of the total deposited stake by certifiers.

\subsubsection{Certifiers}\label{sec:Cert}

To become a certifier, a MC stakeholder needs to create a special transaction  $CertifierReg(SC_{ID}, $ $pubKey) $ on the mainchain that will lock his deposit indefinitely until another special transaction  $CertifierWithdraw$ $(reg_{txid},$  $  sig(pubKey))$ is created.  A single registration will give the certifier rights to be eligible for certificate signing only for a specific sidechain $SC_{ID }$. $pubKey$ is used to sign cross-chain certificates and \textit{CertifierWithdraw }transaction.

To be eligible in a certain epoch $ i$, a certifier needs to create a registration transaction before this epochs begins. 

\textbf{} 

\textbf{Definition (Certifier Group).} \textit{For a determined set of eligible certifiers} ${EC}^{e{pid}}=  $  $\{c_0,...,$ $c_{|{EC}^{e{pid}}-1|}\}$\textit{ in the epoch epid, a certifier group }$CG_i$\textit{, where }$i\in \{0,..,\left[\frac{|{EC}^{e{pid}}|}{N}\right]-1\}$\textit{, will consist of }$N$\textit{ certifiers randomly chosen in }${EC}^{e{pid}}$\textit{ according to the \hyperref[alg:CG_construct]{Algorithm 2}.}

\textit{ The set of eligible certifiers }${EC}^{e{pid}}$\textit{ consists of }${EC}^{e{pid}}\in RG\backslash PC$\textit{, where RG is the set of all registered certifiers, and PC is the set of certifiers that participated in the previous k certifier groups (k is the length of the dispute period).}
\begin{algorithm}[!th]
\label{alg:CG_construct}
\caption{Certifier group construction}
	\begin{algorithmic}[1]
	\STATE \textit{For each certifier }$c_j\in {EC}^{e{pid}}$ \textit{ calculate his personal lottery ticket }${ticket}_{c_j}=$   $H(rand\ ||\ epid\ ||\ $ $pubKey_{c_j})$\textit{, where }$H(\cdot )$\textit{ is a cryptographically strong hash function.}
	
	\STATE \textit{ Sort all lottery tickets in the ascending order.}
	
	\STATE \textit{ Take tickets }$(i\cdot N,\ i\cdot N+1,\ ...\ ,\ i\cdot N+N-1)$\textit{ and include their owners into the certifier group }$CG_i$. 
\end{algorithmic}
\end{algorithm}

Given the above rules up to $g = \frac{|{EC}^{epid}|}{N}$ certifier groups can be deterministically defined for a specific epoch with randomness $rand$\textit{. }It can be seen that a certain certifier cannot be assigned for different groups in the same epoch and also is not eligible during the dispute period after the participation.

To determine the set of certifiers of a specific CCCert, a simple rule is followed: a CCCert with index $ i$ should be signed by the certifier group with the same index. So there is a one-to-one mapping between certificates and certifier groups: ${CCCert_{i} \rightarrow CG_{i}}$. Following these rules, it is easy to identify what should be signed and by whom. A certifier cannot belong to more than one certifier group per withdrawal epoch.

The stages of the backward transfer process are summarized in Fig. 
\ref{fig:11StcerPrc}:

\begin{figure}[htbp]
	\centering
 \includegraphics[trim={0cm 5.53cm 0cm 2.58cm}, clip,width=.86\columnwidth]{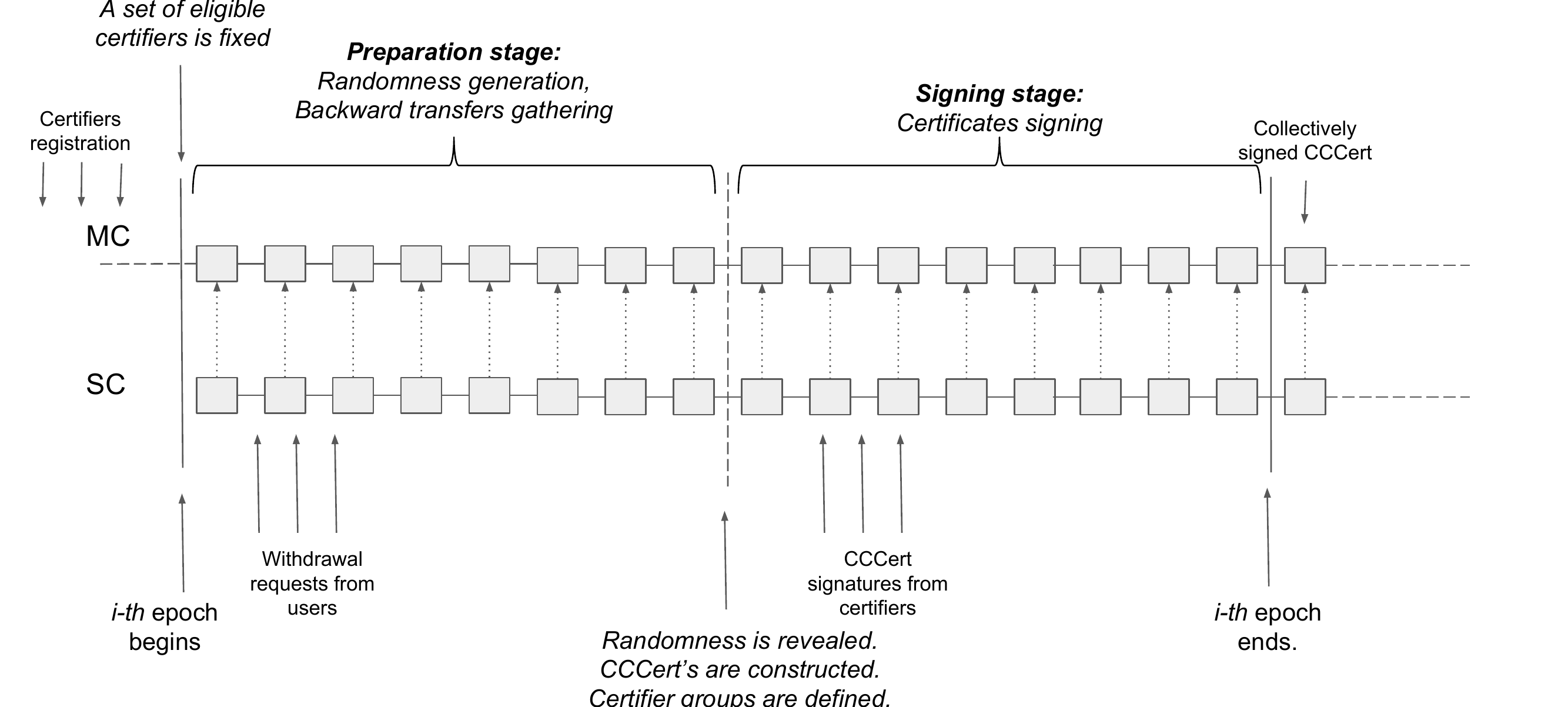}
	\caption{ Stages of certification process}
	\label{fig:11StcerPrc}
\end{figure}

It can be seen that after the preparation stage is finished, the signing process is started and lasts until the end of the epoch. During the signing epoch, each elected certifier should create and submit his signature for the corresponding CCCert. The signatures should be submitted strictly before the epoch ends. After the epoch ends, the signatures for each CCCert (there should be at least $s\ =\left[\frac{N}{2}\right]+1$ signatures, where $ N$ is the size of the certifier group) should be collected into a single aggregated signature and submitted with the corresponding CCCert to the mainchain.

\textbf{Randomness.} For the certifiers' selection procedure, randomness is required. To provide a reliable source of randomness we suppose to use simple yet effective mechanism that leverages the mainchain proof-of-work algorithm. The randomness for a particular epoch $E_{id}$ is equal to the smallest proof hash of the mainchain blocks during the preparation stage of epoch $E_{id}$. In this case, it can be shown that to successfully manipulate randomness, an adversary will need to possess a significant amount of hashing power. One attempt to change the randomness will cost approximately the same as mining the whole preparation stage. We argue that such attack is infeasible and economically unprofitable under reasonable assumptions regarding the honest hashing power ratio.

The formal analysis of such approach is planned for future work as separate research.

Note that it is crucial to fix the set of eligible certifiers before the randomness generation period is started. Once the certifiers' public keys are fixed, the only way to manipulate the selection procedure is to manipulate the randomness itself which, as we mentioned previously, is infeasible under certain assumptions.

\textbf{Incentives.} All selected certifiers from the certifier group $CG_{i}$ are rewarded with $k\%$ (e.g. $k = 1$) of the withdrawal amount of $CCCert_{i}$ distributed equally among certifiers. Only signing certifiers are to be rewarded. Rewards are going to be included by the forger in the SC block at the beginning of the following epoch after the $CCCert_{i}$ has been submitted to the mainchain. Certifiers' reward balance will be managed by SC (MC will not be aware about certifiers reward mechanism).

\begin{figure}[htbp]
	\centering
 \includegraphics[trim={1cm 11.57cm 4cm 2.9cm},clip,width=.86\columnwidth]{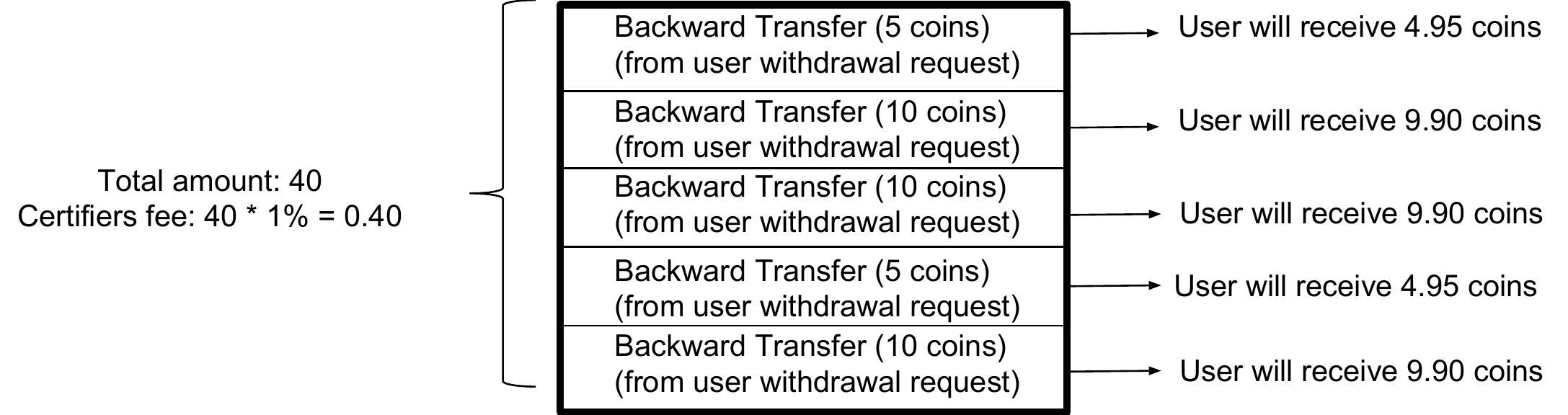} 
 \centering
	\caption{An example of the backward transfer list that contains 5 backward transfers. Transfer amounts are specified with certifiers' rewards already subtracted.
	}
	\label{fig:12Exbactrlst}
\end{figure}

\textit{Fig. \ref{fig:12Exbactrlst}} shows the backward transfers list with all the transfers. For each backward transfer a fee is charged that will be redistributed later to the certifiers who signed the certificate. 

\begin{figure}[htbp]
	\centering
	 \includegraphics[trim={0cm 4.3cm 0cm 4.5cm},clip,width=.86\columnwidth]{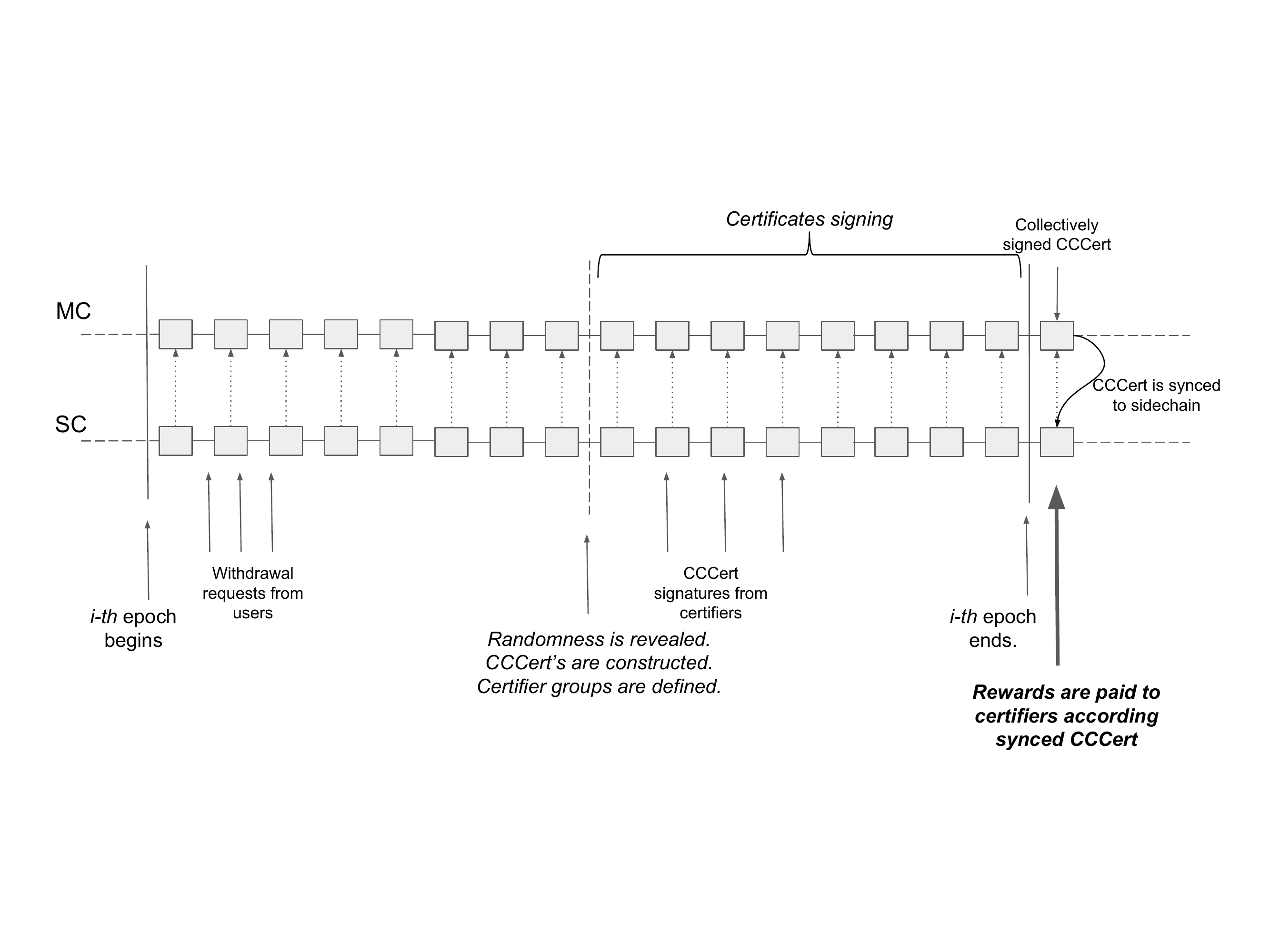}
	\caption{ Rewards to certifiers}
	\label{fig:13Rewcer}
\end{figure}

So, naturally, the reward for the certifiers is a fee paid by those who make the transfer. The reward itself is paid at the beginning of the following epoch at the same sidechain block where the CCCert from mainchain was synchronized, as shown in \textit{Fig. \ref{fig:13Rewcer}.}

The amount of reward is equal for all certifiers. If some of them missed providing a signature, their part of the reward will be burned. The reward is paid only in the case of inclusion of a valid cross-chain certificate that was first submitted to the mainchain.\textbf{}

\textbf{Aggregated signature. }The cross-chain certificate submitted to the mainchain is accompanied with the aggregated signature constructed from the personal signatures of certifiers. Different schemes exist that can be used, e.g. in the simplest form it can be just a concatenation of personal signatures. More advanced schemes are also possible, e.g. BLS threshold signatures. The choice of the appropriate scheme is a matter of further research.

\textbf{Certifier revocation. }To revoke the certifier's rights and unlock the deposit, a certifier needs to create a special transaction $CertifierWithdraw(reg\_tx_{id}, sig).$ The deposit will be unlocked after the dispute period is finished, and only in the case that there is no fraud reported for the certifier (fraud reports and disputes are discussed in Section \ref{sec:WithSaf}).

If the $CertifierWithdraw(reg\_tx_{id}, sig)$ transaction was submitted before the signing stage, the dispute period for the certifier starts from the current epoch. If it was submitted after the signing stage begins, then the certifier is still eligible during this epoch, and the revocation process will be started only in the following withdrawal epoch.

\subsubsection{Withdrawal safeguard}\label{sec:WithSaf}

The \textbf{safeguard} is a special feature introduced to prevent unlimited withdrawals from a sidechain to the mainchain in the case of a total certifiers' corruption. The essence of the safeguard function is to maintain the balance of a sidechain and to restrict withdrawals for amounts larger than were previously transferred to a sidechain.

Implementation of the safeguard feature is simple: for each deployed sidechain, a special balance variable is maintained by the mainchain. Each forward transaction increases the balance by the transferred number of coins. Each withdrawal certificate reduces the balance by the withdrawn amount. The cross-chain certificate cannot withdraw more coins than stored in the sidechain balance.

Such a simple feature prevents any possible implications of a sidechain corruption for the mainchain. It guarantees that only the transferred number of coins can be withdrawn back to the mainchain. Even in the case of total corruption, an adversary cannot print coins on the mainchain out of thin air.

\subsubsection{Disputes}\label{sec:Disp}

To secure the system further a special mechanism of disputes is introduced to prevent signing of a malicious Cross-Chain Certificate. We will call such certificate a \textbf{fraudulent certificate.}

A \textbf{fraudulent certificate} is a certificate that is confirmed in the mainchain but does not correspond to the certificate signed previously in the sidechain. Since the MC does not follow the sidechain and is not able to verify consistency and validity of the certificate directly, a special dispute algorithm is employed to define whether a particular cross-chain certificate is fraudulent. 

The dispute procedure is as described: the following  $k$ consecutive cross-chain certificates (starting from the following withdrawal epoch) can include a Fraud Report signaling a fraudulent certificate. If at least one such report is included in the CCCert during the dispute period, then the certificate under consideration is perceived to be fraudulent, and all signers of this certificate are punished by destroying their deposits.  $k$  is a security parameter and can be tuned upon sidechain creation.

The certifier is unable to withdraw his deposit if some of his certificates (which were signed by him) are considered fraudulent.

Provided that a particular epoch may contain several cross-chain certificates, all of them participate in fraud detection. 

\begin{figure}[htbp]
	\centering
 \includegraphics[trim={0cm 6.5cm 0cm 3cm},clip,width=.8\columnwidth]{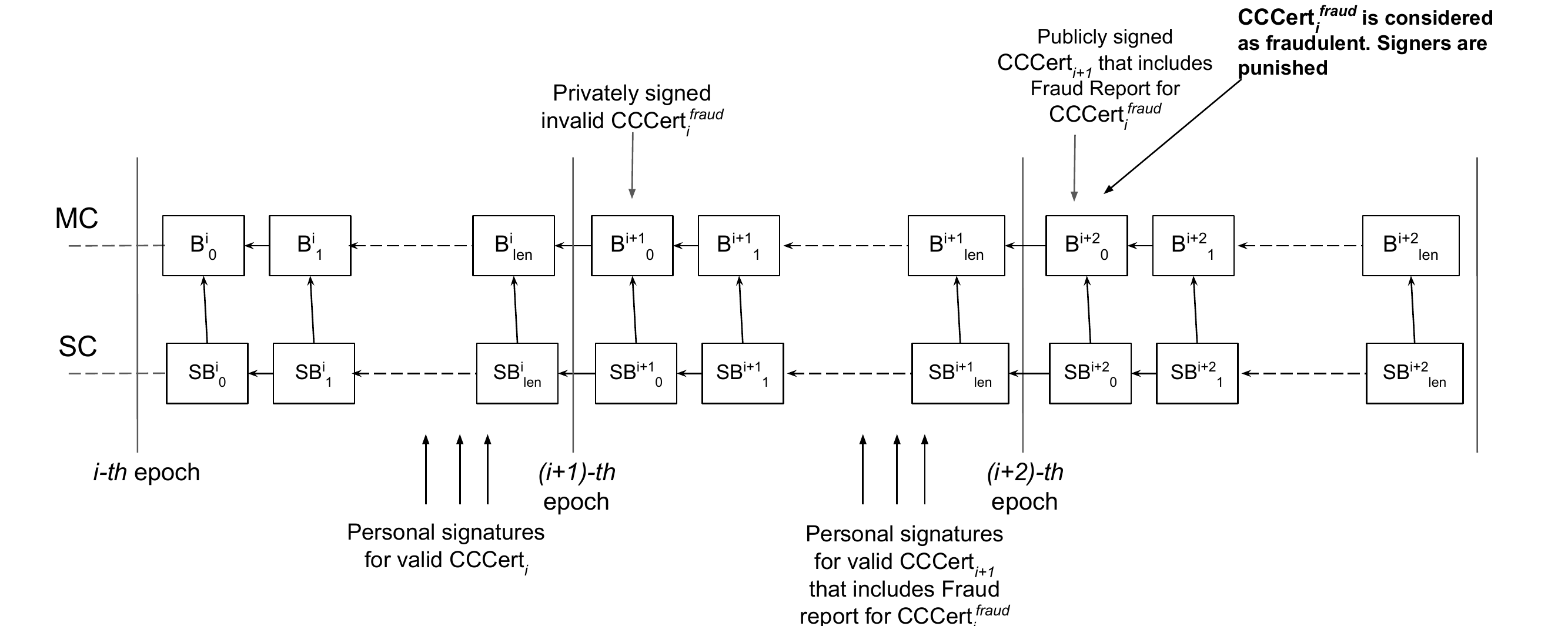}
	\caption{ An example of fraudulent certificate and fraud report}
	\label{fig:14Exfrep}
\end{figure}

The basic fraud reporting flow can be modeled with the \hyperref[algo3]{Algorithm 3} (see also Fig. \ref{fig:14Exfrep}):

\begin{algorithm}[!th]
	\caption{Fraud reporting flow}
	\label{algo3}
	\begin{algorithmic}[1]  
		 \STATE Cross-Chain Certificate  $CCCert_{i}$  is correctly signed in SC by the certifier group  $CG_{i}$ .
		
		\STATE   Fraudulent Certificate  $CCCer{{t}_{i}}^{fraud}$  is created privately, signed by the majority of certifiers of  $CG_{i}$  and then sent to MC.
		
		\STATE    $CCCer{{t}_{i}}^{fraud}$  is included into a block ${B}_{0}^{i}$ in MC. It is valid in the mainchain because it contains a valid aggregated signature. The mainchain does not verify consistency of the certificate in a sidechain.
		
		\STATE    $CCCer{{t}_{i}}^{fraud}$  is synced back in sidechain block ${SB}_{0}^{i}$.
		
		\STATE   The certifiers of the epoch  $i+1$  compare  $CCCer{{t}_{i}}^{fraud}$  with  $CCCert_{i}$ , detect the difference and include a Fraud Report into the  $CCCert_{i+1  }$ referring to the fraudulent certificate  $CCCert_{i}^{fraud}$ 
		
		\STATE   Cross-Chain Certificate  $CCCert_{i+1  }$ is created including the fraud report for  $CCCert_{i}^{fraud}$ 
		
		\STATE   Cross-Chain Certificate  $CCCert_{i+1}$   is signed in SC by the certifier group  $CG_{i+1}$  .
		
		\STATE   Cross-Chain Certificate  $CCCert_{i+1}$   is included into block ${B}_{0}^{i+1}$ in MC.
		
		\STATE   MC will not allow to unlock withdrawals for all the certifiers that previously signed  $CCCert_{i}^{fraud}$ 
	\end{algorithmic}
\end{algorithm}
 
Since fraud detection is a completely deterministic process from the sidechain side, it is required from the honest parties to include a fraud report if the certificate(s) from the $k$ previous withdrawal epochs were fraudulent. Otherwise, it itself becomes fraudulent and should be punished by the following certificate.

\subsubsection{Security}

It can be shown that under rational behaviour assumption the backward transfer protocol is secure against cheating by certifiers. By cheating we imply that a colluding group of certifiers signs a malicious certificate withdrawing coins to their own addresses instead of those defined by the withdrawal requests in the sidechain. The punishment system makes such an attack unprofitable as cheating certifiers would lose their deposits. Since the withdrawal amount per certificate is limited and an adversary needs at least half of the cerfier group to be involved in cheating, it is easy to see that more coins would be confiscated than obtained back.

It also can be shown that punishment of cheating certifiers is practically inevitable under assumption that majority of the all registered certifiers is honest.

The rigorous mathematical analysis of the mentioned security properties is planned to be published in a separate paper.

\subsection{Bootstrapping sidechain}\label{sec:BtstrSid}

There can be many sidechains simultaneously deployed and operating. To simplify and standardize a sidechain deployment procedure, a special sidechain creation transaction needs to be introduced in the mainchain. Such transaction notifies the mainchain about the sidechain creation, sets the id of the sidechain and its different parameters.

The creation transaction has the following structure:
 
\[CreateSidechainTx\ =\ \{ledgerid,start\_block,epoch\_len,\ prep\_len,cert\_depo,\] 
\[cert\_group\_size,cert\_fee,min\_transfer\_amount,dispute\_len\}\]

where

\textbf{\textit{ledgerid}} is a unique identifier of the sidechain that hasn't yet been registered;

\textbf{\textit{start\_block}} is the block number on the mainchain from which the first withdrawal epoch begins;

\textbf{\textit{epoch\_len}} is the length of the withdrawal epoch in blocks;

\textbf{\textit{prep\_len}} is the length of the preparation stage of the withdrawal epoch;

\textbf{\textit{cert\_depo}} is the needed amount of the locked deposit to become a certifier;

\textbf{\textit{cert\_group\_size}} is the size of the Certifier Group that certifies a single cross-ledger certificate;

\textbf{\textit{cert\_fee}} is the fee percentage from the transferred amount that will be paid to certifiers;

\textbf{\textit{min\_transfer\_amount}} is the minimum transfer amount allowed for a single backward transfer;

\textbf{\textit{dispute\_len}} is the number of cross-ledger certificates that can report on fraud of a particular  $CCCert_{i}$  (starting from the following withdrawal epoch after the epoch where  $CCCert_{i}$  occurred).

Customizable parameters give flexibility in choosing of suitable parameters for a particular sidechain. Different combinations of the provided parameters give different trade-offs among speed, efficiency, and security.

Using this protocol, an Horizen sidechain can be created by anyone. A deployment fee (in a form of burned coins) should be applied by the sidechain creator.

Once a sidechain is created, the withdrawal epochs are deterministically defined and forward/backward transfers can be processed. Note that even if some withdrawal epochs do not have cross-chain certificates (because of absence of certifiers or any other reason), it does not disrupt the mainchain processing flow and interaction with a sidechain may still happen in future. Even if the sidechain was suspended for whatever reason, the withdrawal still may happen in future under the defined schedule.
	\section{  Conclusion}
	
	In this paper we presented the sidechain construction based on proof-of-stake principles. Our main contribution is the construction of the new backward transfers protocol that relies neither on a centralized federation of validators nor miners/block forgers. Instead, it relies on a new set of actors in the system called ``certifiers''. We also provided a description of the consensus protocol that can be used in the sidechain and fits well to its structure. Our construction requires alteration of the mainchain consensus protocol to integrate basic sidechain support. Once integrated, many sidechains can be deployed. The mainchain needs to track only the balance and the set of current certifiers for a particular sidechain. In general, the proposed construction provides a decentralized open protocol for a sidechain support. It can be integrated into both proof-of-work and proof-of-stake blockchains.

\printbibliography

	\end{document}